# Photonic Crystal Spatial Filters Fabricated by Femtosecond Pulsed Bessel Beam

DARIUS GAILEVIČIUS,[1,2*] VYTAUTAS PURLYS,[1,2] AND KESTUTIS STALIUNAS[3,4]

[1] *Vilnius University, Faculty of Physics, Laser Research Center, Sauletekio Ave. 10, Vilnius, Lithuania*
[2] *"Femtika", Saulėtekio al. 15, LT-10224, Vilnius, Lithuania*
[3] *Institució Catalana de Recerca i Estudis Avancats (ICREA), Passeig Lluís Companys 23, 08010, Barcelona, Spain*
[4] *Universitat Politècnica de Catalunya (UPC), Rambla Sant Nebridi 22, 08222, Terrassa (Barcelona) Spain*
*\*darius.gailevicius@ff.vu.lt*

**Abstract:** We propose and experimentally demonstrate femtosecond direct laser writing with Bessel beams for the fabrication of photonic crystals with spatial filtering functionality. Such filters are mechanically stable, of small (of order of millimeter) size, do not require direct access to the far-field domain, and therefore are excellent candidates for intracavity spatial filtering applications in mini- and micro-lasers. The technique allows the fabrication of efficient photonic crystal spatial filters in glass, with a narrow angle (~1 degree) nearly 100%-transmission pass-band between broad angle (up to 10 degrees) nearly 0%-transmission angular stop-bands. We show, that this technique can not only significantly shorten the fabrication time, but also allows the fabrication of large-scale defect-free photonic crystal spatial filters with a wide filtering band.

## 1. Introduction

Spatial filtering is a general technique used to improve the spatial properties of the radiation of lasers. Typically, conventional lasers (e.g. solid-state lasers) contain low angle pass intracavity spatial filters in form of confocal arrangement of lenses with the circular diaphragm in the confocal plane. The diaphragm ensures the operation of the laser on the lowest possible order transverse mode [1]. However, such an arrangement is very inconvenient or even impossible in microlasers, such as microchip lasers, edge emitting semiconductor lasers, or VCSEL lasers. The millimeter-order length of the resonators does not allow the conventional confocal lens arrangement for the intracavity spatial filtering. This is possibly the main reason, why the radiation from many of microlasers suffers from the bad spatial beam quality, which is especially problematic in higher emission power regimes. Lifting the limitations of the beam spatial quality would allow to increase the brightness of the radiation of microlasers and would open new relevant areas of their technological application.

A promising solution of the beam spatial quality problem is the intracavity use of compact Photonic Crystal (PhC) spatial filters. The idea of PhC spatial filtering was proposed in [2,3], and subsequently experimentally demonstrated in [4–6] in different realizations. The idea is based on a selective diffraction of the angular components of the light propagating through the double-periodic photonic structure: the particular angular components of the incident light at a resonance with the transverse and longitudinal periodicities of the photonic structure diffract efficiently and are removed from the zero-diffraction order of the transmitted beam. The PhC spatial filtering, together with the super-collimation [7], already showed itself as a powerful tool to clean the spatial structure of the beams, which could be efficiently employed in microresonators, for instance in the microchip [8] and diode [9] lasers. For a detailed review on PhC spatial filtering see e.g. [10].

In practice, the fabricated PhC filters, especially those written by femtosecond pulses into inorganic materials using direct laser writing techniques, have a problem of a limited functionality of spatial filtering. Firstly, the rough estimations show [10], that the depth of filtering band scales as $\Delta n \cdot l/\lambda$, where $\Delta n$ is the refraction index modification (which in glasses is of the order of $10^{-3} - 10^{-2}$), $l$ is the length of the written photonic crystal and $\lambda$ is the wavelength. This means that the efficient filtering, with 100% depth of the filtering dips can be achieved for >100 $\mu m$ length of the PhC. Moreover, the angular width of the filtering line is proportional to $\Delta n$, which is the fractions of degree for such low-index-contrast PhCs. For relevant applications the filtering line of several degrees is required (for instance the main technological challenge in edge emitting semiconductor lasers is to reduce the slow-axis divergence from ~ 5 degrees down to 1-2 degrees). The width of the filtering band is not a fundamental problem, since it can be substantially broadened using chirped photonic crystals [6,11], i.e. by sweeping the geometrical parameters along the photonic structure, which moves the relatively narrow filtering band through the broader area of angular transmission spectrum. Such sweep of the filtering band could cover larger angular ranges, see Fig.1(a) for the illustration, however, eventually would result in respectively longer crystals.

Thumb rule [10] says that the 100% filtering of the angular range $\Delta \varphi$ relates with the length $l$ of the chirped photonic crystal as $l = \lambda \Delta \varphi / \Delta n$. For example, if $\Delta n$ is small, of order of $10^{-2}$ and if the desired angular range of filtering is $\Delta \varphi \sim 0.1$ in radians, then one needs a crystal length of the order of 1 mm. Typically the PhCs are fabricated by direct laser writing using Gaussian beams, where tightly focused femtosecond laser beam induces local modification of refractive index in glass or other transparent materials [12–15]. The convenient transverse periodicity of the PhCs is in the range of 1-2 µm, which means that a relatively high numerical aperture (NA) focusing optics must be used (NA > 0.8), however, this contradicts with the required length of the PhC (> 1 mm). The existing variety of focusing optics forces to make a choice between high NA, but short working distance optics, and long working distance, but low NA optics. In addition, the deeper in the glass substrate is the focus, the more it is affected by the spherical aberrations. In practice, spherical aberrations and limited working distance limit the length of PhCs to appr. 0.3 mm.

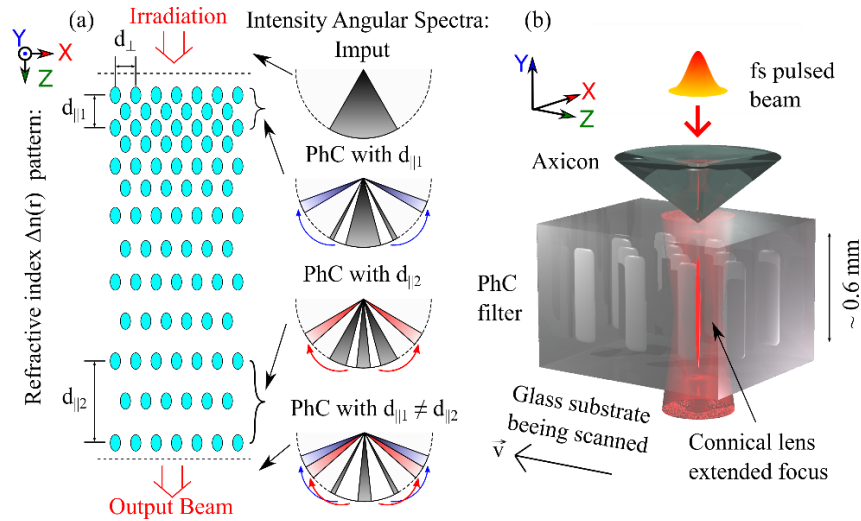

Fig. 1. (a) Illustration of broad band spatial filtering in chirped PhCs. (b) Principle scheme of direct laser writing using Bessel beams, illuminated in vertical Y-direction, whereas the filtering follows along the horizontal Z-direction.

For some applications the filtering along one direction is sufficient. For example, it is especially suitable for broad edge emitting semiconductor lasers, where only the slow-axis filtering is required [16,17]). In such case the PhCs of 2D geometry are needed. The idea presented in this paper is to use the Bessel beams [18] for fabrication of such large longitudinal scale 2D PhC spatial filters.

Bessel beams can be generated by refractive, reflective or diffractive axicons and are characterized by a long and narrow high aspect ratio high intensity focusing area. This property is used for machining transparent materials [19–21], e.g. for fabrication of phase elements [22,23]. Due to very long and narrow area of refractive index modifications, the fabrication can be performed in a "horizontal geometry", see Fig.1(b). Then the length of the filtering structure is no longer a problem and in principle is limited only by the mechanical travel range of the fabrication stages. Although the vertical direction is still limited, it is at least by one order of magnitude longer compared to that using the Gaussian writing beam strategy.

The main message of the article is an experimental demonstration of a possibility to inscribe sufficiently long PhC crystals by using Bessel beams, which can ensure technologically relevant 1D spatial filtering. We show the efficient filters in the visible and in the near infrared range. The letter describes: Bessel beam writing setup; fabrication of the real structure; experimental observation of the light angular transmission characteristics through such PhC structures of different length.

## 2. Fabrication and characterization

The samples were fabricated using a direct laser writing setup, which is schematically shown in Fig. 2(a). The Bessel beam was generated using a UVFS (n ≈ 1.45) conical lens (axicon) with a half angle of ∠α = 0.5° (cor. apex angle: 179°). The incoming beam was of a Gaussian profile with a diameter of $2w = 5.3$ mm at $e^{-2}$ level. The center wavelength $\lambda = 1030$ nm, the pulse repetition rate 25 kHz and pulse duration approximately 200 fs. Such configuration led to an extended focal line (Bessel zone) length of approximately $Z_{BZ1} \approx 1.34$ m. In order to decrease its dimensions a demagnifying telescope was arranged from two lenses with the focal lengths of 500 mm and 9 mm, having a demagnification value of M~55.6. Pulse energy delivered to the sample corresponded to 8 µJ.

A polished N-BK7 ($n_{ref} \approx 1.51$) 4 mm thick rectangular substrate was used as a substrate and was mounted on a 3D positioning stage. After demagnification inside the substrate the Bessel zone length $Z_{BZ2}$ was estimated to be around 600 µm (according Snell's law as in [24]). The sample was scanned at 2500 µm/s in a consecutive linear motion to produce the patterns shown in Fig. 2(b). The usable aperture of such PhCs is $0.6 \times 1$ mm$^2$ and can be extended in the horizontal (filtering) direction without limitation. We note that the usable aperture in the y-direction as shown in Fig. 2(b) is less than the estimated Bessel zone length $Z_{BZ2}$. Practically, $Z_{BZ2}$ was expected to be longer since we did not use a spatial filter to correct the distortion by the axicon tip [25–27]. The oscillatory refractive index modifications seen at the bottom of the facet view are a result of this. For the exposure conditions used here, positive refractive index changes are induced during fabrication according to [28,29].

After fabrication we inspected the PhC samples with a transmission optical microscope, Fig. 2(b) and then probed them illuminating by focused Gaussian laser beam. We observed the angular filtering bands as shown in Fig. 2(c). To record the transmitted intensity pattern, we projected the beam directly on a CCD camera matrix placed at the far-field.

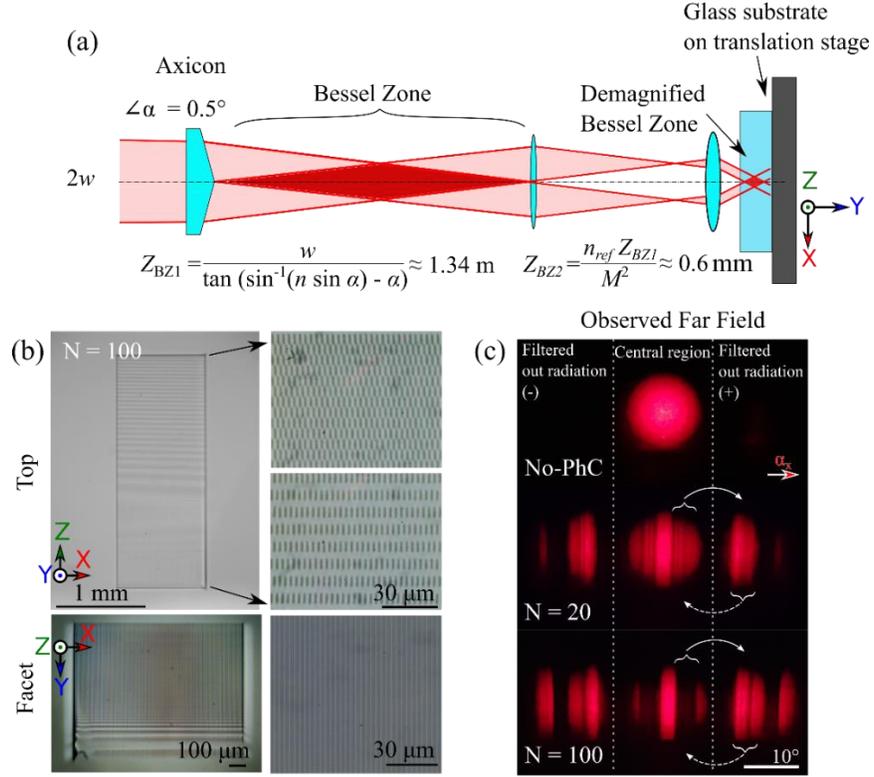

Fig. 2. Illustration of the fabrication arrangement (a), microscope-photos of the sample (b), and the 2D light transmission picture, showing the central beam region, from which radiation is filtered out (c). Arrows indicate resonant energy transfer between two coupled angular bands: outward and, also, inward to the central region.

The described fabrication scheme allowed us to achieve a minimum transverse period of 3 µm, accordingly by a ~1.5 µm transverse width of the inscribed voxels. The central filtering angle of non-chirped PhCs for a wavelength $\lambda = \lambda_{vac}\, n_{ref}$ in paraxial approximation reads: $\sin(\alpha_c) = \lambda(Q-1)/2d_\perp$, where $Q$ is a geometry parameter, $Q = 2d_\perp^2/(\lambda d_\parallel)$, chosen according to the target filtering angle. For example, in order to have a central filtering angle of $\alpha_c = 1$ deg for $\lambda_{vac} = 633$ nm wavelength and $d_\perp = 3$ µm transverse period, the $Q$ is ~1.25 ($d_\parallel$ ~34.4 µm respectively). For chirped PhCs the $Q$ factor is varied along the structure. E.g. if the $Q$ is varied from 1.2 to 1.4, the central filtering angle varies from 0.8 to 1.6 degrees.

### 3. Results

To explore the spatial filtering in regular (non-chirped) PhCs we fabricated several samples with different numbers of longitudinal periods $N$ ($d_\parallel \approx 26.8$ µm). The target filtering wavelength was $\lambda_{HeNe} = 633$ nm. Experimentally measured transmission spectra are shown in Fig. 3(a). The filtering dips in the angular transmission function are getting deeper with increasing the PhC length until they reach the maximum depth at $N = 8$ periods. If the length of the PhC is further increased, the filtering regions become shallower again. This is due to the fact that the filtered radiation is propagating along the crystal couple back to the original modes. At $N = 14$ (approximately double the optimum length for filtering) almost no filtering regions are observed. These are so called Laue-Rabi oscillations, see [10] for a more detailed explanation, with the revival period of $N \approx 14$.

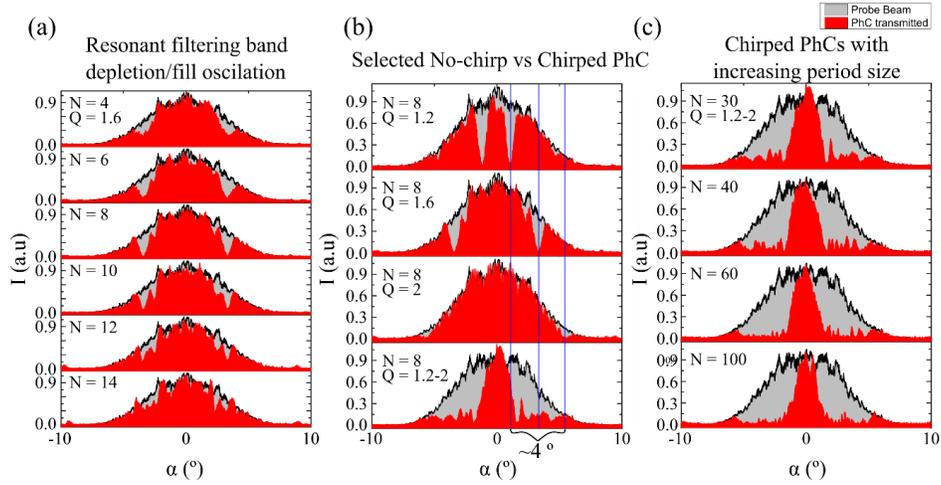

Fig. 3. Transmitted probe beam angular intensity spectra for different PhCs without (a,b) and with chirp (b,c) and for different crystal lengths (a,c) illuminated by $\lambda_{HeNe} = 633$ nm. (c) shows transmitted beams for increased PhC lengths for chirped case, where saturation of the filtering effect occurs.

Next, we selected several $Q$ values $Q$ = 1.2, 1.6, 2.0 (estimated central filtering angles: $\alpha_1$= 0.8°, $\alpha_2$ = 2.4°, $\alpha_3$ = 4°) and fabricated three PhC samples with fixed longitudinal period and optimal length ($N$ = 8 periods) in order to demonstrate the filtering angle dependence on the longitudinal period. The angular transmission spectra, Fig. 3(b), show that the filtering angles correspond reasonably well to the estimated values and were measured to be $\alpha_1$= 1.08°, $\alpha_2$ = 3.2°, $\alpha_3$ = 5.5°. The discrepancy is due to paraxial approximation used to derive the angle *vs.* $Q$ relation. The width of the filtering bands in all three cases was about 0.6°.

In order to increase the width of the filtering band a chirped PhC geometry was used (1.2 ≤ $Q$ ≤ 2). The $d_\parallel$ values were varied linearly in a range 21.5 ≤ $d_\parallel$ ≤ 35.8 µm. The bottom graph of Fig. 3(b). shows the corresponding angular transmission spectrum of such structure of $N$ = 30 periods. The positive effect of chirping is obvious when comparing with Fig. 3(b) column. Although fabricated with the same conditions, the chirped PhCs exhibit much wider filtering band of $\Delta\alpha \approx 4°$, compared to $\Delta\alpha \approx 0.6°$ for non-chirped examples. The filtered angular bands are a bit shallower for $N$ = 30 periods chirped crystals compared to the $N$ = 8 periods non-chirped cases, however this can be corrected by increasing the number of periods as it is shown in Fig 3(c).

The geometry of the PhC can be tuned in order to match the target wavelength. In Fig. 4(a) the results for $\lambda_{IR} = 970$ nm and the same $Q$ interval as previously are shown. In this case the longitudinal period was varied in 14 ≤ $d_\parallel$ ≤ 23.35 µm range. Here the filtering angle range is extended even further. We observed filtering for 2.2° ≤ |α| ≤ 8.05°, corresponding to the angular bandwidth of $\Delta\alpha \approx 5.8°$. In addition, we quantified the removed energy and the total incident energy ratio (Fig. 4(b)). As the PhC length (the number of longitudinal periods $N$) increases, the filtering performances increases also, and, opposite to the non-chirped case, saturates to the 100% filtering performance.

For comparison, the numerical simulations of the spatial filtering through the PhCs with varying refraction index contrast $\Delta n$ were also performed and presented in Fig. 4(b). The simulations were performed using a beam propagation method approach, where a complex amplitude scattering efficiency parameter $s$ was used [30] (the physical meaning is that for a plane-wave at resonant angle the $s^2$ represents the intensity scattered into one diffraction component per longitudinal half-period). The saturation to 100% filtering for smaller $\Delta n$ occurs for longer PhCs, as expected. The simulations with $s = 0.14$ correspond well with experimental measurement data. Using this value, we determined the refraction index contrast in the inscribed PhCs: $\Delta n \approx 2\lambda_0 s/\pi^{3/2} l_0 \approx 5 \cdot 10^{-3}$ ($l_0 \approx 6.9$ µm is the refractive index modified region length).

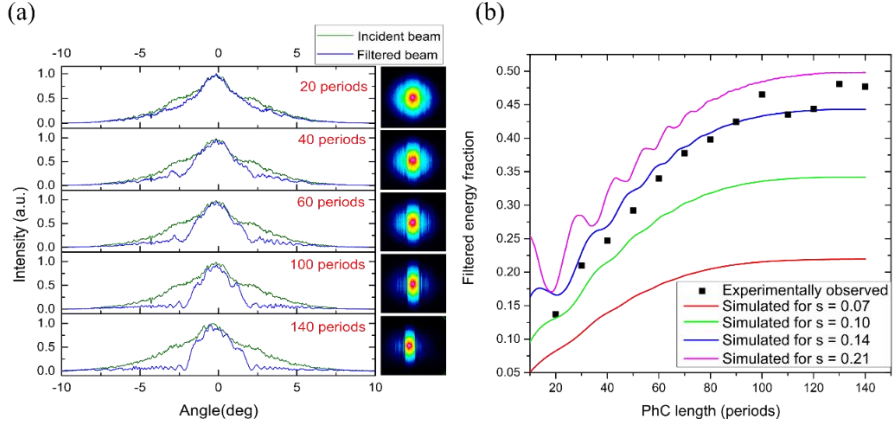

Fig. 4. (a) Experimentally measured angular transmission intensity spectra of PhCs designed for λ = 970 nm operating wavelength. Spatial filtering band is seen at 2.2° ≤ |α| ≤ 8.05°. (b) The corresponding data (black dots) for the change in total transmitted energy opposed to the PhC length and compared with four numeric cases (solid curves).

Judging from the examples in Fig. 3(c) and Fig. 4 the chirped PhCs provide a clear advantage over the non-chirped ones when a broad filtering band is required. The filtering band width can in principle be further extended by fabricating longer crystals using the described technique.

The total length of the filtering PhC structures was in the range of a few millimeters. For example, the PhC structures designed for both 633 nm and 970 nm wavelengths at N = 30 periods were of ~2 mm length. Such relatively high resolution ($d_\perp = 3$ µm) and more importantly long structures are challenging to fabricate with the Gaussian beam point-by-point direct laser writing method because of the limited working distance and/or optical aberrations in the sample. Also, in practice their fabrication duration is relatively long. Let us make a simplified comparison. A PhC of an aperture of 600×600 µm² and 16 periods and Q = 1.2 takes around 3 minutes to produce using the Bessel technique. A PhC having the same aperture and transverse period using the reported scanning speed of v = 2.5 mm/s and a raster scan pattern would take around 30 minutes [6].

## 4. Conclusions

In this way we proposed new, efficient method for inscribing relatively long photonic crystal spatial filters. This technique, however, optimally works for the 2D photonic crystals, providing 1D spatial filtering, for the use, for instance, in broad area semiconductor lasers, where the slow-axis filtering is highly desired. The divergence of the beams along the slow axis is a real problem in such lasers, where typically, the divergences reach 3 to 10 degrees. The technique proposed by us provides the angular filtering range, sufficient for substantial improvement of the spatial structure in typical broad are semiconductor lasers.

Advanced chirping schemes could be applied to make use of the large number of periods to either extent the filtering range or to tune the filter angular transmission spectrum shape. Making a vertically stitched structure could increase the usable aperture by a factor of 2-5 times.

## 5. Funding, acknowledgments, and disclosures

*5.1 Funding*

This work was supported by the EUROSTARS Project E!10524 HIP-Lasers, as well as by Spanish Ministerio de Ciencia e Innovación, and European Union FEDER through project FIS2015-65998-C2-1-P. D.G. and V.P. acknowledge the financial support from "FOKRILAS" Project S-MIP-17-109 from Research Council of Lithuania.

*5.2 Acknowledgments*

The authors are grateful to Optician Šarūnas Jablonskas from Laser Research Center for preparing the BK7 samples.

*5.3 Disclosures*

The authors declare no conflicts of interest.